# CONCAVE MICROLENS ARRAY MOLD FABRICATION IN PHOTORESIST USING UV PROXIMITY PRINTING

*Tsung-Hung Lin[1], Hsiharng Yang[2,3] and Ching-Kong Chao[1]*

[1] Department of Mechanical Engineering, National Taiwan University of Science and Technology, Taipei, Taiwan 106
[2] Institute of Precision Engineering, National Chung Hsing University, Taichung, Taiwan 402
[3] Center of Nanoscience and Nanotechnology, National Chung Hsing University, Taichung, Taiwan 402

## ABSTRACT

This paper presents a simple and effective method to fabricate a polydimethyl-siloxane (PDMS) microlens array with a high fill factor, which utilizes the UV proximity printing and photoresist replication methods. The concave microlens array mold was made using a printing gap in lithography process, which utilizes optical diffraction of UV light to deflect away from the aperture edges and produces a certain exposure in the photoresist material outside the aperture edges. This method can precisely control the geometric profile of concave microlens array. The experimental results showed that the concave micro-lens array in photoresist could be formed automatically when the printing gap ranged from 240 μm to 720 μm. High fill factor microlens array can be produced, when the control pitch distance between the adjacent apertures of the concave microlens array was decreased to the aperture size.

## 1. INTRODUCTION

Microlens arrays have extensive applications in optical communication and optical interconnection. The primitive microlens array may originate from several fabrication methods. The photoresist reflow method [1], the UV proximity printing method [2], and grayscale mask [3], focused-ion-beam (FIB) milling [4], and etching [5] to make the master of microlens array. Electroforming process can be applied to replicate the microlens array mold in metal. A great quantity of polymeric microlens arrays can be fabricated using micro-compression, injection molding, or hot embossing by using the electroformed mold inserts of microlens array. Furthermore, polymeric microlens arrays are applied to the charge-coupled-device (CCD) cameras [6], optical scanners, fiber interconnects [7, 8, 9], copiers [10], liquid-crystal displays (LCDs), and personal digital accessories (PDAs) applications.

Polydimethyl-siloxane (PDMS) was used in a soft lithography technique proposed as a replication material [11]. K. Watanabe et al. developed a three-dimensional microlens mold made by focused-ion-beam chemical vapor deposition [12]. The microlens mold was used to fabricate PDMS microlens array. Also, the PDMS microlens with variable focal lengths was developed by liquid-filling in a cavity [13]. The PDMS microlens array was also applied to a pneumatic actuator [14]. PDMS microlens arrays used a replication molding technique was fabricated. To fabricate the PDMS microlens using molding, a master microlens array as the same profile as the final PDMS microlens array is needed. The photoresist reflow method is used to make the master microlens array. The photoresist reflow method is to melt photoresist structures to form microlens array due to the surface tension. After completing the master microlens array in photoresist, the next step is to replicate the microlens array in photoresist to a PDMS mold by using casting method. The PDMS mold is cured in a vacuum oven for 2 hrs at 5 mTorr of pressure at 75℃, the mold is peeled off from the master microlens array. The PDMS mold consists of a concave microlens array. The final step is to spin-coating PDMS on the PDMS mold. Depending on the spinning speed, a thick or thin film with the desired dimensions of microlens array can be obtained. Then, PDMS film is cured for 2 hrs at 5 mTorr of pressure at 75 ℃. Finally the PDMS microlens array was peeled of the PDMS mold.

The lens shape and the fill-factor are two of many conditions that impact to overall light efficiency. To collect the maximum amount of light, the lens area must be as close to 100% of the total area as possible [15]. The fill-factor is defined as the percentage of lens area to the total area. And the thermal reflow processing conditions must be delicately controlled to create a high fill-factor microlens array [16], which increases the process difficulties and reproducibility requirements. In this study, the UV proximity printing followed by a photoresist mold is proposed here to provide a simple method of fabricating PDMS microlens arrays. A PDMS microlens array with an extremely high fill-factor was produced.

## 2. EXPOSURE CHARACTERISTICS

Lithographic exposure has three printing modes including contact, proximity, and projection. The exposure operations using the proximity mode is in the near field or Fresnel diffraction regime. The pattern resulting from the light passing through the mask directly impacts onto the photoresist surface because there is no lens between the photoresist and mask. The created aerial image therefore depends on the near field diffraction pattern. Because of the diffraction effects, the light bends away from the aperture edges and produces partial exposure outside the aperture edges. Although the contact mode can minimize these effects by reducing the gap to zero, the gap is not strictly zero in practice because the top surface of the photoresist is not perfectly flat.

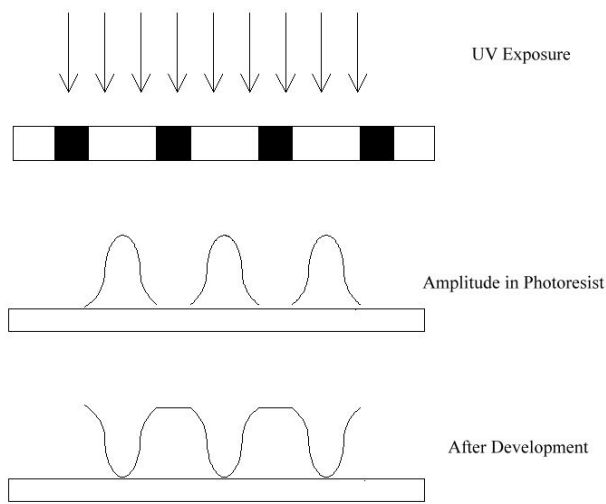

Figure1. Light intensity distribution in positive photoresist exposed using adjacent aperture

Figure 1 shows the intensity distribution in the photoresist and after development as the light passes through the apertures. The middle figure shows a smooth and convex light intensity profile in photoresist. The concave profile can be fabricated through proper operational parameters using a positive photoresist and the desired microstructures are formed after the development process as shown in the bottom figure. As the gap spacing increases between the mask and photoresist, the aerial image quality produced on the photoresist surface will strongly degrade due to the diffraction effects [17]. That is, the intensity distribution range becomes wider and the relative light intensity becomes weaker at some point for one aperture exposure. The aperture exposure arranged using the array produces a complicated three-dimensional light intensity distribution in the photoresist. The final concave mold of microstructure geometry can be determined using the resulting intensity distribution after exposure and development.

### 3. PDMS Microlens fabrication

Figure 2 shows the proposed PDMS microlens fabrication processes. Desired patterns are transferred from the designed mask in the lithographic process. In this experiment, a plastic mask was fabricated using a laser writing onto a PET (Polyethylene terephthalate) used for PCBs (print circuit boards). The layout pattern on the plastic mask is illustrated in Figure 3. Each aperture was d=80μm and the pitch distance for two adjacent apertures was p=120μm. The upper and lower rows were arranged in equidistance. Round patterns laid out in an ortho-triangle on the PET-based mask.

Figure 2(a) sketches the lithography process to transfer patterns onto the substrate using the proximity printing. A gap was controlled between the mask and the photoresist surface. A silicon wafer was used as the substrate. To increase the adhesive force between the photoresist and substrate, the silicon wafer substrate was firstly coated with a thin layer of HMDS. The substrate was then spun with a positive photoresist (AZ4620) layer 18μm thick. The spin condition was 500rpm for 40 seconds. Prebaking in a convection oven at 90℃ for 3 minutes is a required procedure. This removes the excess solvent from the photoresist and produces a slightly hardened photoresist surface. The mask was not stuck onto the substrate. The sample was exposed through the plastic mask using a UV mask aligner (EVG620). This aligner had soft, hard contact or proximity exposure modes with NUV (near ultra-violet) wavelength 350-450nm and lamp power range from 200-500 W. A slice of glass was inserted between the photoresist and mask to create a gap. Exposure was then conducted for about 8 seconds. The three-dimensional array was completed after exposure and dip into the developer for 2 minutes and cleaning with deionized water as shown in Figure 2(b). A liquid poly-dimethylsiloxane (PDMS) was cast on the photoresist pattern indicated in Figure 2(c) and peeled off from the photoresist mold after it was cured shown in Figure 2(d).

The curing condition is 2 hrs at 75℃. The printing gap was adjusted to 120, 240, 360, 480, 600, 720, and 840 μm for finding different results. The PDMS microlens variations in the samples were observed and the optimal gap range determined. Optical microscopy (OM), scanning electron microscopy (SEM) and NanoFocus μScan 3D laser profilometer were used to measure the characteristics of the resulting PDMS microlens structures.

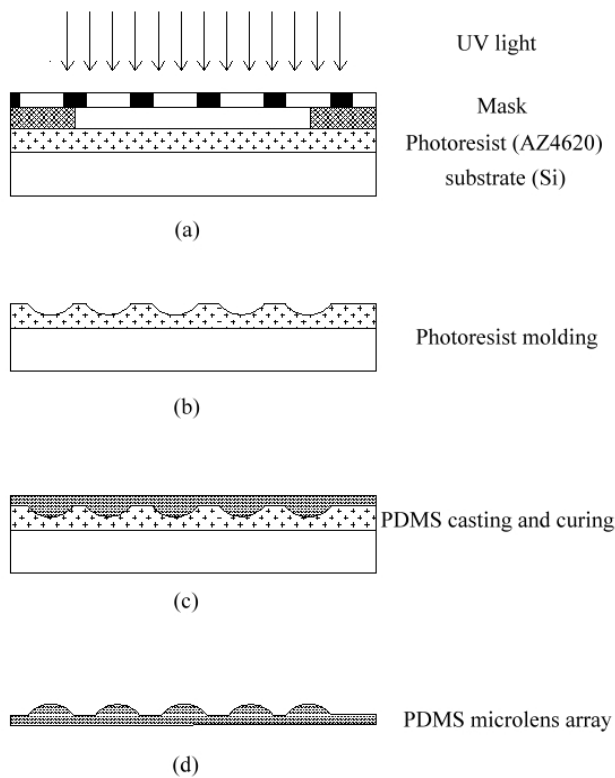

Figure 2. Flowchart of PDMS microlens fabrication process; (a) UV proximity printing, (b) Development, (c) PDMS casting and curing, (d) Peel off of PDMS microlens array.

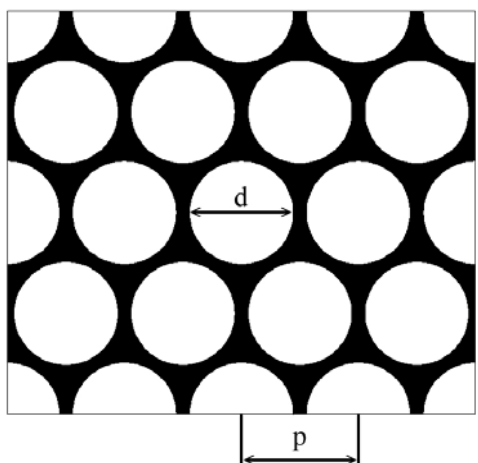

Figure 3. Illustration of the pattern design on the mask for concave microlens array mold fabrication, the upper and lower rows of equidistance were designed. (Unit: μm)

## 4. RESULTS AND DISCUSSION

### 4.1 Printing gap size effect

According to the experimental results, three PDMS microlens array mold were classified after exposure by using different gap sizes. The printing gap sizes were ranged from 120 μm to 840 μm. Firstly, a flat down microlens array mold was formed using the printing gap 120 μm. Figure 4 (a) shows the flat top structural array mold by using OM observation. Secondly, adjusting the printing gap between 240μm to 720μm, microlens arrays mold with concave hemisphere were formed as shown in Figure 4 (b). Thirdly, using the printing gap is larger than 720 μm, the top view is blurred as observed using OM shown in Figure 4(c). The printing gap between the photoresist and mask gradually increased, the height and sternness in the PDMS microlens array mold at the lateral sideline became thinner and flatter. A small printing gap (less than 240 μm) is not suitable for PDMS microlens array mold fabrication because the light intensity distribution in the photoresist is not have enough diffraction to produce concave spherical structures.

A SEM microlens array mold in photoresist produced using a 360μm printing gap is shown in Figure 5. The concave microlens surface is quite smooth. A liquid poly-dimethylsiloxane (PDMS) was cast on the concave microlens array mold and peeled off from the photoresist mold after it was cured. The PDMS microlens array was fabricated. The SEM micrograph of the PDMS microlens surface using a 360μm printing gap is shown in Figure 6. Its actual profile compared to the theoretical spherical curve (dashed line) of a spherical lens is depicted in Figure 7. It shows that the deviation on the lens sag from a perfect sphere is quite small.

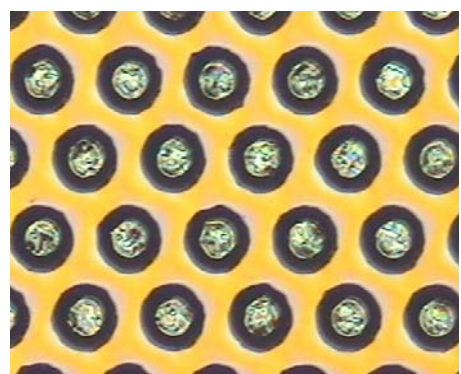
(a)

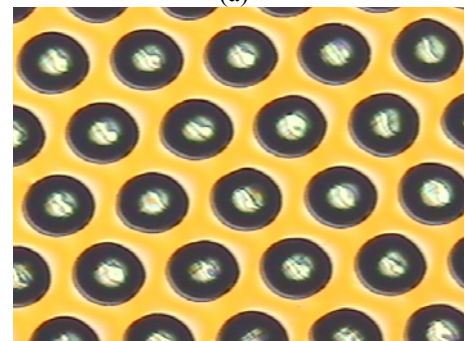
(b)

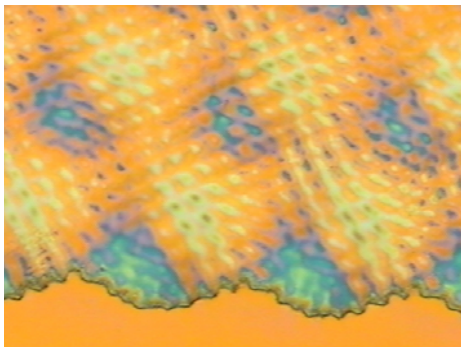

(c)

Figure 4. OM photographs of microlens array mold in photoresist using different printing gap sizes; (a) 120, (b) 360, and (c) 840μm

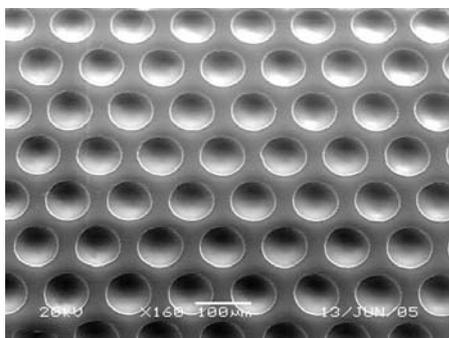

Figure 5. SEM micrograph of the microlens mold fabricated using a 360 μm printing gap.

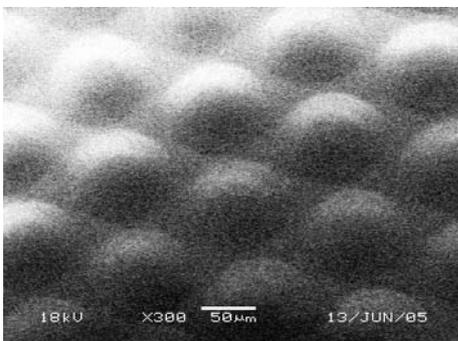

Figure 6. SEM micrograph of microlens array in PDMS fabricated using a 360 μm printing gap.

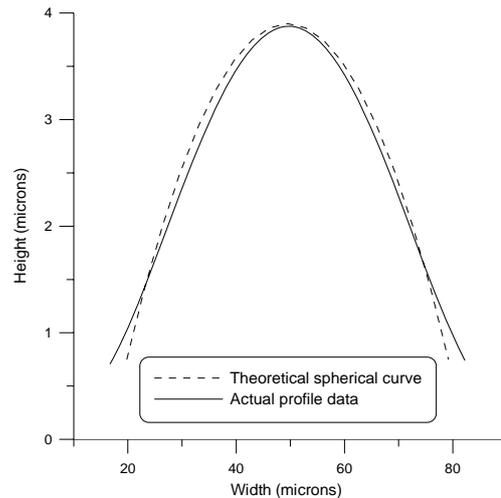

Figure 7. Cross-section profile of a microlens as measured with a stylus profilometer.

4.2 Pitch distance effect

To fabricate the microlens array mold in photoresist by using a 360 μm printing gap was confirmed. Different pitch distances for two adjacent apertures were designed onto the mask. Four different pitch distance patterns including 90, 100, 110, and 120 μm were designed. Four patterns of concave microlens arrays in photoresist are observed using OM as shown in Figure 8. A SEM photograph showing an extremely high fill factor of microlens array mold using the pitch distance 90 μm is shown in Figure 9. Pitch distance close to the aperture size can result a high fill factor microlens array. The 3D profile of microlens array in PDMS and its cross sectional view is shown in Figure 10. It was measured using a NanoFocus μScan 3D laser profilometer. The lens sag height is 4.03 μm in Figure 10 (b). The bottom edge of every microlens was adjacent to each other. Gapless microlens array in PDMS is achieved. The microlens array in PDMS is actually with extremely high fill factor.

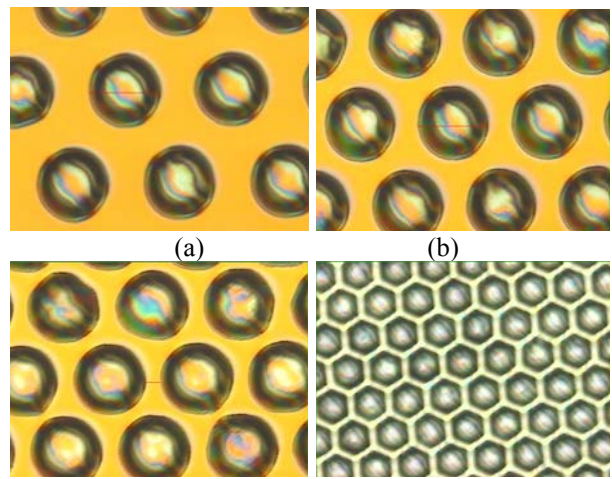

(a)   (b)

(c)                  (d)

Figure 8. OM photographs of concave microlens array mold in photoresist using different pitch distances; (a) 120, (b) 110, (c) 100, and (d) 90 μm.

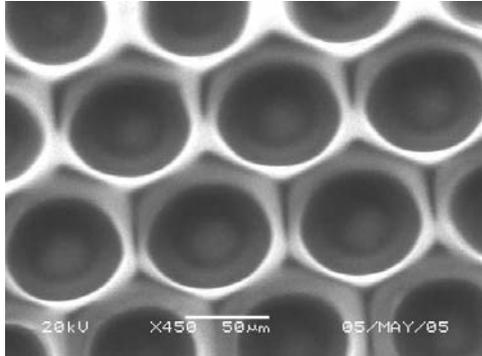

Figure 9. SEM micrograph of high fill factor of microlens array mold fabricated using a 10μm pitch distance for two adjacent apertures and a 360μm printing gap

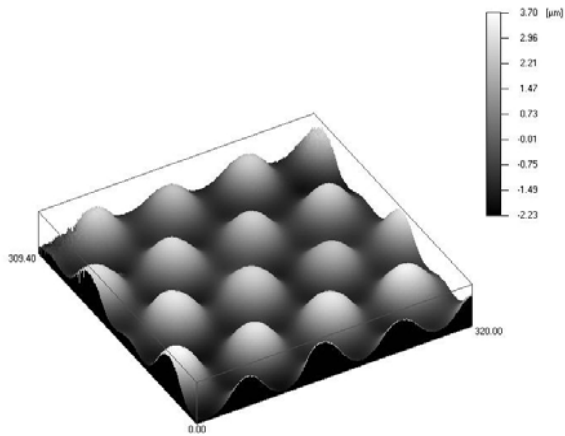

(a)

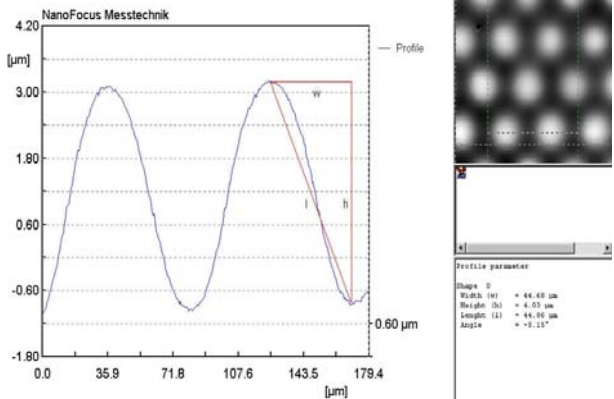

(b)

Figure 10. Microlens array in PDMS measured by NanoFocus μScan 3D laser profilometer; (a) 3-D surface profile and (b) the cross-sectional profile of PDMS microlens array.

### 4.3 Microlens array profiles

Using the proximity printing method to fabricate concave microlens array in photoresist and furthermore to replicate convex microlens array in PDMS is practical. The convex PDMS microlens array geometry formed is summarized in Table 1. *D* is the PDMS microlens diameter, *f* is the focal length and *h* is the sag in the PDMS microlens. The data are based on the experimental results and calculated using the following method. According to the basic geometrical consideration [18], the radius of curvature (*RC*) was calculated by Equation (1). The focal length was calculated using Equation (2). The microlens focal length was obtained using the PDMS with a refractive index (*n*) 1.44 [14]. The numerical aperture (*NA*) was calculated using Equation (3). The PDMS microlens array surface roughness was determined using an atomic force microscope (AFM). The measured size was 5×5μm$^2$ on the PDMS microlens top. It shows that the microlens surface roughness (Ra) was 2.31n m in Figure 11.

$$Rc = \frac{h^2 + \frac{D^2}{4}}{2h} \quad (1)$$

$$f = \frac{Rc}{n-1} \quad (2)$$

$$NA = \frac{D}{2f} \quad (3)$$

Table 1. Summarized geometrical data of microlens array in PDMS fabricated by using the proximity printing method.

| gap (μm) | D (μm) | h (μm) | RC (μm) | f (μm) | NA |
|---|---|---|---|---|---|
| 240 | 111.8 | 8.58 | 186.39 | 423.61 | 0.1320 |
| 360 | 116.16 | 6.11 | 279.10 | 634.32 | 0.0916 |
| 480 | 116.26 | 5.97 | 285.99 | 649.98 | 0.0894 |
| 600 | 118.46 | 5.92 | 299.26 | 680.14 | 0.0871 |
| 720 | 122.5 | 4.9 | 385.26 | 875.60 | 0.0700 |

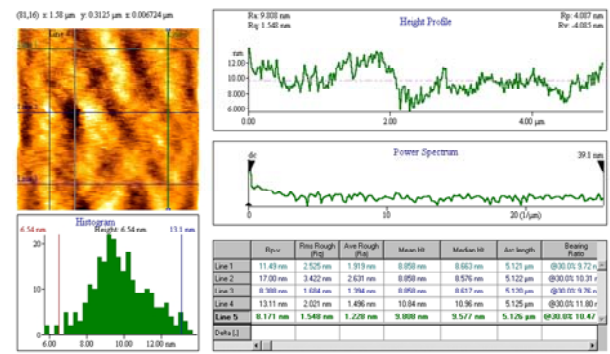

Figure 11. Surface roughness of PDMS microlens measured by AFM

## 5. CONCLUSION

A different approach to fabricate convex microlens array in PDMS is presented. The convex PDMS microlens array with a fill factor of almost 100% can be fabricated by using the proximity printing and replication method. Two important parameters are summarized. The printing gap sizes are twice or more than the pitch distance and the pitch distance are close to the mask aperture size. The proposed method has advantages over previous fabrication methods. The fabrication process is very simple. Printing gaps ranged from 240 to 720 μm (twice or more than the pitch distance) using the same pattern on the mask can generate concave microlens in photoresist. The PDMS microlens array mold with a fill factor of almost 100% can be fabricated in a lithographical process.

## 6. ACKNOWLEDGEMENT

This work was supported by the National Science Council (series no. NSC94-2212-E-005-016) of Taiwan, R.O.C.

## 7. REFERENCES

[1] Z. D. Popovic, R. A. Sprague and G. A. Neville Connell "Technique for monolithic fabrication of microlens arrays," *Appl. Opt.*, pp. 1281-97, 1988.

[2] C.-P. Lin, H. Yang, and C.-K. Chao, "A new microlens array fabrication method using UV proximity printing," *Journal of micromechanics and micro-engineering*, pp. 748-757, 2003.

[3] L. W. Cheng, "Fabrication and analysis of diffractive optical elements," *Thesis of Department of Physics*, National Central University, pp. 51-52, 2001.

[4] Y. Fu and N. K. Annbryan, "Semiconductor microlens fabricated by one-step focused ion beam direct writing," *IEEE Tran. On Semiconductor Manufacturing*, Vol. 15, No. 2, pp. 229-231, 2002.

[5] M. B. Stern and T. R. Jay, "Dry etching for coherent refractive microlens arrays," *Optical Engineering*, Vol. 33, No. 11, pp.3547-3551, 1994.

[6] K. Itakura, T. Nobusada, N. Kokusenya, R. Nagayoshi and M. Ozaki, "A 1-mm 50k-pixel IT CCD image sensor for miniature camera system," *IEEE Tran. on Electron Devices*, Vol. 47, No. 1, pp. 65-70, 2000.

[7] J. S. Leggatt and M. C. Hutley, "Microlens arrays for interconnection of single-mode fiber arrays," *Electron. Lett.*, Vol. 27, pp. 238-240, 1991.

[8] Y. Fu, N. K. Annbryan, and O. N. Shing, "Integrated micro-cylindrical lens with laser diode for single-mode fiber coupling," *IEEE Photon. Technol. Lett.*, Vol. 12, No. 9, pp. 1213-1215, 2000.

[9] S. R. Cho, J. Kim, K. S. Oh, S. K. Yang, J. M. Baek, D. H. Jang, T. I. Kim, and H. Jeon, "Enhanced optical coupling performance in an nGaAs photodiode integrated with wet-etched microlens," *IEEE 122 Photon. Technol. Lett.*, Vol. 14, No. 3 pp. 378-380, 2002.

[10] G. P. Smith, "Some recent advances in glasses and glass-ceramics," *Materials & Design*, Vol. 10, No. 2, pp. 54-63, 1989.

[11] Y. Xia and G. M. Whitesides, "Soft Lithography," *Angewandte Chemie International Edition*, Vol. 37, pp. 550-575, 1998.

[12] K. Watanabe, T. Morita and R. Kometani, "Nanoimprint using three-dimensional microlens mold made by focused-ion-beam chemical vapor deposition," *Journal of Vacuum Science and Technology B: Microelectronics and Nanometer Structures*, Vol. 22, p 22-26, 2004.

[13] J. Chen, W. Wang, J. Fang, and K. Varahramyan, "Variable-focusing microlens with microfluidic chip," J*ournal of Micromechanics and Microengineering*, pp. 675-680, 2004.

[14] K. Hoshino and I. Shimoyama, "An elastic thin-film micro-lens array with a pneumatic actuator," *IEEE*, pp. 321-324, 2001.

[15] N F Borrelli," Efficiency of microlens array for projection," *LCD 44th Electronic Components and Technology Conf.*, pp. 338–345, 1994.

[16] H. Yang, C.-K. Chao, M. K. Wei and C.-P. Lin, "High fill-factor microlens array mold insert fabrication using a thermal reflow process," *Journal of micromechanics and micro-engineering,* pp. 1197-1204, 2004.

[17] J. D. Plummer, Silicon VLSI Technology, Prentice Hall, pp. 208-234, 2000.

[18] S. Sinzinger and J. Jahns, *Microoptics*, Wiley-VCH Verlag GmbH, Weinheim, Germany, pp. 6-10, 1999.